\documentstyle[prl,aps]{revtex}
\input epsf

\tighten

\begin{document}


\title{Estimate of the primordial magnetic field helicity}

\author{
Tanmay Vachaspati}
\address
{
Department of Astronomy and Astrophysics,
Tata Institute of Fundamental Research,\\
Homi Bhabha Road, 
Colaba, Mumbai 400005, India\\
and\\
Department of Physics,
Case Western Reserve University,\\
10900 Euclid Avenue,
Cleveland, OH 44106-7079, USA.}

\wideabs{
\maketitle

\begin{abstract}
\widetext
Electroweak baryogenesis proceeds via changes in the non-Abelian 
Chern-Simons number. It is argued that these changes generate 
a primordial magnetic field with left-handed helicity. The helicity 
density of the primordial magnetic field today is then estimated 
to be given by $\sim 10^2 n_b$ where $n_b\sim 10^{-6}$ /cm$^3$ is 
the present cosmological baryon number density. With certain 
assumptions about the inverse
cascade we find that the field strength at recombination 
is $\sim 10^{-13}$ G on a comoving coherence scale $\sim 0.1$ pc.
\end{abstract}
}

\narrowtext

The helicity of a primordial magnetic field is of crucial
importance in determining its subsequent evolution and in
assessing whether the observed magnetic fields can be generated
by amplification of the seed field by a galactic dynamo
\cite{ZelRuzSok83}. The average helicity density of a magnetic 
field, {\bf B}, in a chosen volume $V$ is defined as:
\begin{equation}
{\cal H} = {1\over V} \int_V d^3 x ~ {\bf A}\cdot {\bf B} 
\label{helicitydefn}
\end{equation}
where ${\bf B}={\bf \nabla}\times {\bf A}$.

The connection between the helicity of primordial magnetic
fields and baryon number is arrived at by considering the
process of electroweak baryogenesis occurring at the time of
the electroweak phase transition \cite{Cor97}. The genesis of baryons 
requires changes in the Chern-Simons number
\begin{eqnarray}
CS = {{N_F} \over {32\pi^2}}
\int d^3 x \epsilon_{ijk} \biggl [ 
 g^2 \biggl ( W^a _{ij} W^{a k} &-& {g\over 3} W^{ai}W^{bj}W^{ck} \biggr)
 \nonumber \\
       &-& {g'}{}^2 Y^{ij} Y^k \biggr ] \ .
\label{chernsimons}
\end{eqnarray}
where, $N_F$ is the number of particle families, $W^{\mu a}$,
$Y^{\mu}$ are the SU(2) and U(1) hypercharge gauge fields,
$g$ and $g'$ are SU(2) and U(1) gauge couplings, and 
$i,j,k =1,2,3$.
Changes in $CS$ are achieved via the production and dissipation
of non-perturbative field configurations such as the electroweak
sphaleron \cite{Man83}, or linked loops of electroweak 
string\cite{VacFie94,GarVac95}, or other equivalent configurations 
(for a review of electroweak strings see \cite{AchVac00}). 
In the cosmological setting, it is believed that such configurations 
would be produced in the false vacuum phase due to the detailed 
dynamics of the electroweak phase transition, and would then decay 
in the true vacuum phase. Once 
they decay, the baryon number so produced cannot be washed out 
by the subsequent production and dissipation of more sphalerons or 
equivalent configurations.

A feature of the baryon number producing intermediary field 
configurations is that they carry fluxes of non-Abelian magnetic 
fields that are twisted or linked. The simplest configuration to 
analyze is that of two linked electroweak strings (Fig. 1). Here
the loops carry {\bf Z}-magnetic flux and the {\bf Z} magnetic 
lines are linked with each other.

Now consider the decay of the linked {\bf Z}-string configuration.
One path to decay is by the breaking of the loops of string. This 
occurs by the creation of monopole-antimonopole pairs. The segments 
of string then collapse. If the loops are very large, the whole 
process would be embedded in the highly conductive cosmological 
plasma in which the electromagnetic magnetic fields are frozen-in. 
Then the electromagnetic magnetic ({\bf B}) field lines remain linked 
and the final helicity of the {\bf B} field is related to that of 
the original {\bf Z}-strings.

\begin{figure}
\epsfxsize =  0.9 \hsize  \epsfbox{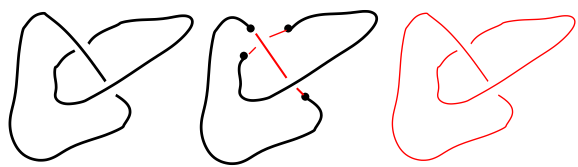}
\vskip 0.5 truecm
\caption{\label{loops}
In the first figure, a pair of linked loops of {\bf Z}-string are
shown. They decay by the nucleation of monopole-antimonopole
pairs (black circles in the middle figure) which get pulled
apart by the strings. Magnetic field ({\bf B}) lines (grey curves)
run between the monopole and antimonopole. The whole system
is embedded in a perfectly conducting plasma in which the
{\bf B} field lines are frozen-in. Finally the strings have decayed,
leaving behind a {\bf B} field configuration with helicity.
}
\end{figure}

Other decay channels of the strings seem possible. For example,
the loops could collapse before the strings break. 
In general we expect that the strings will collapse and decay into
modes of {\bf B} -- since this is the only massless gauge field
in the theory -- and that these modes
will retain some of the original helicity of the {\bf Z} field.
For an order of magnitude estimation of the helicity, it is sufficient
to assume that the decay channel described above is not exceedingly 
improbable\footnote{Numerical simulations of electroweak strings 
do show finite segments of strings that evolve by the motion of 
the monopoles at the ends \cite{AchBorLid99}. Also, for Abelian 
fields in vacuum, an initially helical state asymptotically evolves 
to a configuration with half the initial helicity \cite{JacPix99}.}.

An electroweak sphaleron may also be interpreted as a segment of 
string terminating on a monopole and an antimonopole 
\cite{VacFie94,HinJam94,Vac94,Hin94}. The helicity is non-vanishing 
because the {\bf Z}-string has a twist. The sphaleron is unstable to the 
untwisting of the monopole with respect to the antimonopole. This will 
transfer helicity from the twisted {\bf Z}-string
to the {\bf B} field lines going from the 
monopole to the antimonopole as the sphaleron decays. 



Having established a connection between the genesis of baryons
and the helicity of magnetic fields, it is straight-forward to
relate one to the other. The de-linking process shown in
Fig. 1 leads to a change in the baryon number by 
$N_F \cos (2\theta_w)$, where $\theta_w$ is the weak mixing 
angle (see \cite{VacFie94,GarVac95}), if the initial 
{\bf Z}-string configuration has linkage $-1$. 
The {\bf B} field flux of an electroweak
monopole is $\Phi_A = (4\pi /e) \sin^2\theta_w$, where $e$
is the electromagnetic coupling (see \cite{Nam77,AchVac00}). 
The helicity of the final 
configuration in Fig. 1 is $- 2 \Phi_A^2$. Hence, every baryon 
that is produced, causes a change in electromagnetic helicity 
\begin{equation}
\Delta {H}  = - 2 \left ( {{4\pi}\over e} \right ) ^2 
                  {{\sin ^4 (\theta_w )}\over {N_F\cos (2\theta_w)}}
                \sim - 100
\label{deltaH}
\end{equation}
where we have used $\sin^2 \theta_w =0.23$, $N_F=3$, and 
$e^2/4\pi \equiv \alpha = 1/137$.
Note that only changes in baryon number are related to changes in helicity.
We further assume that the initial helicity density is negligible and 
this allows us to estimate the final helicity as $\Delta H$. (We
are assuming that there are no sources besides electroweak
baryogenesis for generating magnetic helicity in the very early 
universe.) The observed number density of baryons is 
$\sim 10^{-6}$/cm$^3$ \cite{Pee93}. Therefore the average helicity 
of the primordial magnetic field is estimated to be 
$\sim - 10^{-4}$/cm$^3$ \footnote{This is precisely the value
of the magnetic helicity assumed in \cite{Cor97}.}. 
In astrophysical units this is $\sim - (10^{-21} {\rm G})^2$-kpc.
Note that this is
not a root-mean-squared value for the magnetic helicity density but 
a mean local value. This is because our universe is observed to be 
made of matter and essentially no antimatter -- supposedly due to
the CP violation present in particle physics. 


Once a helical magnetic field is produced there are a number
of circumstances under which its helicity is conserved. In the 
MHD approximation, the magnetic field in Minkowski spacetime obeys 
the equation
\begin{equation}
{{\partial {\bf B}} \over {\partial t}}
           = {\bf \nabla} \times ({\bf v} \times {\bf B} )
               + {1\over {4\pi\sigma_c}} {\bf \nabla}^2 {\bf B}
\label{Bevoln}
\end{equation}
where ${\bf v}$ is the fluid velocity and $\sigma_c$ is the electrical
conductivity of the plasma. First, if $\sigma_c$ is infinite,
dissipation can be ignored and the evolution of the magnetic field 
only depends on the first term on the right-hand side of
eq. (\ref{Bevoln}).  In this case, even the local helicity -- 
defined by restricting $V$ in the integral in 
eq. (\ref{helicitydefn}) 
to volumes bounded by field lines \cite{Tay74} -- is conserved. 
In a cosmological setting, the dissipation term can be ignored as 
compared to the effects of Hubble expansion on length scales above 
a certain critical scale called the ``frozen-in'' scale: 
$L_f \sim \sqrt{t/ {4\pi \sigma_c}}$,
where $t$ is the cosmic time. Magnetic fields that are coherent on 
scales larger than $L_f$ are said to be frozen-in and their helicity 
is conserved both locally and globally. 

If the fluid velocities are small, it is possible that the first
term on the right-hand side in eq. (\ref{Bevoln}) can be ignored 
and only the dissipation term
is relevant. In this case the field dies out exponentially fast and
helicity is not conserved. However, if the fluid velocity is not
negligible, the evolution of the field depends on both terms.
In this situation, Taylor \cite{Tay74}
has argued that the local helicity of the field changes due to 
reconnections of the field lines and hence is not conserved; however,
the global helicity which is a sum over a lot of random local changes
is still conserved. While there is no proof of Taylor's conjecture,
it leads to a successful explanation of the ``reversed
field pinch''. We shall assume that Taylor's conjecture is true.
The criterion for deciding if global helicity is conserved is that the
magnetic Reynolds number ${\cal R}_M = 4\pi \sigma_c L v$ for magnetic
fields on a length scale $L$ and characteristic fluid velocity
$v$, should be large. With 
$\sigma_c \sim T/\alpha {\rm ln} (1/\alpha ) \sim T/e^2$ 
\cite{TurWid88,BayHei97} and $L \sim 1/e^2 T$,
the length scale characteristic of gauge field configurations
such as the sphaleron, the condition ${\cal R}_M >> 1$ 
gives $v >> e^4/4\pi$. Whether this
condition is met depends on the fluid dynamics during baryogenesis.
Successful electroweak baryogenesis requires significant departures
from thermal equilibrium which is likely to be accompanied by 
large fluid velocities. Therefore we will assume that the fluid
velocities are large enough for ${\cal R}_M >> 1$. As discussed above, 
under these circumstances, the field will evolve while conserving 
global magnetic helicity even though the field is not frozen-in. 

The evolution of the field after production is a central problem
in MHD. There are several features that are expected on
theoretical grounds \cite{Tay74,Son99,FieCar00} and on the basis of 
numerical simulations \cite{BisMul99,Chretal00}. The first feature
is that the magnetic field tangle is expected to evolve towards
a configuration with ${\bf \nabla}\times {\bf B} = \beta {\bf B}$ 
for some constant $\beta$. Provided the field is characterized by
a single length scale, this state is one of ``maximal helicity''
-- a state in which the energy is minimum subject to the constraint 
of fixed global helicity. This expectation is supported by Taylor's 
analysis of the reversed field pinch.
The second feature is that helical magnetic fields are expected to
``inverse cascade'' -- that is, energy will be transferred from
small length scales to large length scales 
\cite{Frietal75,Pouetal76,Son99,FieCar00,BisMul99,Chretal00}
(though also see \cite{BerHoc01}). If $L(t)$ is the coherence
scale of the field, the existing studies at large ${\cal R}_M$ 
in Minkowski spacetime find
\begin{equation}
L(t) = L_* \left ( {t \over {t_*}} \right ) ^\xi
\label{Lscaling}
\end{equation}
where the exponent $\xi$ has been determined to be $1/2$ in numerical
studies \cite{BisMul99,Chretal00}, and $2/3$ in analytical studies 
under various approximations.
The factor $L_*$ is the initial coherence scale of the magnetic
field and $t_*$ is the time scale associated with
the turbulence. We will take $t_* = L_*/v_*$ where $v_*$ is the typical
fluid velocity. The third feature, as decribed earlier, is that 
the evolution is expected to conserve global helicity.

A little care is needed in applying eq. (\ref{Lscaling}) to cosmology.
The reason is that the origin of $t$ in eq. (\ref{Lscaling}) is the
time at the start of the simulation ($t=0$) while in cosmology the
magnetic field is produced at the electroweak epoch. Furthermore,
in cosmology, the flat (non-expanding) MHD equations can be used
provided the time used is the conformal time \cite{BraEnqOle96,SubBar98}. 
In a radiation dominated universe, the conformal time $\tau$ is 
related to the cosmic time $t$ by: 
$\tau = 2t_{ew}^{1/2}(t^{1/2} - t_{ew}^{1/2})$. The factors
of $t_{ew}$ have been chosen to give $\tau =0$ at $t=t_{ew}$
and $a(t_{ew})=1$ where $a(t)$ is the cosmological scale factor. 
Hence, in cosmology, the factor $(t/t_*)^\xi$ in eq. (\ref{Lscaling}) 
should be replaced by $(\tau /t_*)^\xi$.
It is more transparent and approximately equivalent to 
directly use eq. (\ref{Lscaling}) for the first Hubble time -- 
during which the expansion of the universe
is irrelevant to the evolution of the magnetic field -- and then 
use eq. (\ref{Lscaling}) with $\xi$ replaced by $\xi/2$, together
with a comoving factor, for further evolution in the radiation 
dominated epoch. 

We can now evolve the magnetic field from the
electroweak phase transition to the present epoch. We focus on
the coherence scale of the field since the field strength can
then be estimated quite easily using the conservation of helicity.
At the electroweak scale we have seen that the magnetic Reynolds number
is large and hence the coherence length will grow as in 
eq. (\ref{Lscaling}) with $t_* = L_*/v_*$ and $L_* \sim 1/e^2 T_{ew}$.
We find that in one Hubble time, 
$$
L(t_{ew}) \sim {1\over {T_{ew}}} 
                       ( v_* t_{ew} T_{ew}  )^\xi
              \sim {1\over {T_{ew}}} \left (
                       {{T_{P}} \over {T_{ew}}} \right )^\xi
$$
where $T_P = 10^{19}$ GeV is the Planck energy and we have
assumed fluid velocities: $v_* \sim 1$. For the 
slower estimate of the inverse cascade ($\xi =1/2$)
this gives $L_{ew} \sim 10^8 /T_{ew} \sim 10^{-8} t_{ew}$ which 
is greater than the
frozen-in scale ($\sim 10^7/T_{ew}$). For the faster
inverse cascade ($\xi =2/3$), the result is 
$L_{ew} \sim 10^{11}/T_{ew}$. Hence the magnetic field 
becomes coherent on a scale larger than the frozen-in scale within 
one Hubble expansion after production at the electroweak epoch.
In this period we also expect the field to evolve towards maximal
helicity with the dissipation of the non-helical component 
\cite{Jedetal98}. Once the field is maximally helical, further
dissipation does not occur because such fields are force-free
\cite{FieCar00}.

Further evolution of the magnetic field is a combination of Hubble 
expansion and inverse cascade. In a radiation dominated
universe, this gives \cite{FieCar00}
$
L(t) = 
       L(t_{ew}) ( {{T_{ew}}/ T} )^{1+\xi}
$
We also check that the frozen-in scale $L_f$
grows as $(T_{ew}/T)^{3/2}$. Hence the coherence scale remains 
greater than the frozen-in scale for both values of $\xi$
and the helicity of the field continues to be conserved.

The next significant event occurs at the epoch of $e^+e^-$ 
annihilation at $T \sim 0.1$ MeV
since the electrical conductivity of the plasma drops very suddenly.
The electrical conductivity prior to the epoch is given by
$\sigma_c \sim T/e^2$ \cite{TurWid88,BayHei97}
while after this epoch it is given by
$\sigma_c \sim 10^{-10} m_e /e^2$. This corresponds to a
drop in conductivity by a factor of $10^{-9}$ and an increase of
the frozen-in scale by $\sim 10^4$. The coherence
scale just prior to $e^+e^-$ annihilation is:
\begin{equation}
L(t_{ee}-) = L_f (t_{ee}-) {{L(t_{ew})}\over {L_f(t_{ew})}}
                \left ( {{T_{ew}}\over {T_{ee}}} \right )^{\xi-1/2}
\label{LLf}
\end{equation}
where we have denoted quantities at $e^+e^-$ annihilation by
the subscript $ee$ and $t_{ee}-$ denotes the time just prior
to annihilation.
Inserting numbers gives $L(t_{ee}-)=10 L_f (t_{ee}-)$ for
$\xi=1/2$ and $L(t_{ee}-)=10^4 L_f (t_{ee}-)$ for $\xi=2/3$.
Furthermore, at this stage the Reynolds number is 
${\cal R}_M = 4\pi \sigma_c L v = (L/L_f)^2 >> 1$ where we have 
made use of the definition of $L_f$ 
and have estimated the fluid velocity as due to cosmological 
expansion at the scale $L$: $v \sim L/t$. After $e^+e^-$ 
annihilation, however, $L/L_f \sim 10^{-3}$ if $\xi=1/2$ and 
$L/L_f \sim 1$ if 
$\xi=2/3$. In the first case, the magnetic Reynolds number
is small and the fields are coherent on scales smaller than the
frozen-in scale. Therefore with $\xi=1/2$ we expect the field
to get dissipated. Only the Fourier modes of the magnetic field
larger than the frozen-in scale can survive. In the second case, 
the magnetic Reynolds number may or may not be significantly 
larger than one and the coherence length of the field is comparable 
to the frozen-in scale. Therefore the coherence scale of the field 
will grow with the Hubble expansion and there may or may not be 
further inverse cascade. From now on we will only consider the
value $\xi =2/3$ and give estimates of the coherence scale of 
the field both with and without inverse cascade in the post
$e^+e^-$ annihilation universe.

The coherence scale at the recombination epoch can now be estimated 
as the scale at the electroweak epoch multiplied by the corresponding
Hubble expansion factor and the inverse cascade factor:
$$
L_{rec} = L_{ew} \left ( {{T_{ew}}\over {T_{eq}}} \right )^{1+\xi}
        \left ( {{T_{eq}} \over {T_{rec}}} \right )^{1+\xi /2}
$$
where $L_{ew} \sim 10^{-5}$ cms, $\xi =0$ without inverse cascade
and $\xi =2/3$ with inverse cascade, and the last factor takes into 
account the evolution in the matter-dominated era from the epoch of
matter-radiation equality ($T_{eq}\sim 1$ eV) to the epoch
of recombination ($T_{rec}\sim 0.1$ eV). This gives 
$L_{rec} \sim 10^7$ cms without inverse cascade and 
$L_{rec} \sim 10^{14}$ cms with inverse cascade. 

The strength of the field can be estimated by using the conservation
of helicity. At recombination the magnitude of helicity density is 
given by $\sim 10^2 n_{b, rec} \sim 10^{5}$/cm$^3$ or 
$\sim (10^{-13} {\rm G})^2$-($10^{14}$ cms). 
(The baryon density at recombination 
is $z_{rec}^3$ higher than that today where $z_{rec} \sim 10^3$ is the
cosmological redshift at recombination.) So the maximally helical,
primordial magnetic field with {\em comoving} 
coherence scale $\sim 0.1$ pc
has a field strength $\sim 10^{-13}$ G at recombination.

The present scenario also allows us to estimate the net helicity
of the galactic magnetic field. Both magnetic helicity and baryon
number are conserved during galaxy formation when the protogalactic cloud
collapses. During this collapse, the helicity and baryon densities
increase due to the decrease in the cloud volume ($V$ in
eq. (\ref{helicitydefn})) but the ratio of the helicity density to the
baryon density remains unchanged.  Hence the helicity density
of the galactic magnetic field can be estimated to be 
$\sim - 10^2 n_{b,gal} \sim - 10$/cm$^3$ which is 
$\sim - (10^{-19} {\rm G})^2$-kpc. Subsequent evolution of the
magnetic field during galaxy formation, including the
amplification by a turbulent dynamo, does not significantly
change the global helicity \cite{Jix99} and we expect this estimate
to hold even today.  Furthermore, we have earlier noted that the magnetic 
helicity density has the same sign everywhere and is negative. 
Hence the magnetic field in all the different galaxies should be 
left-handed\footnote{This simple prediction is complicated by 
galactic processes that might generate local helicity while conserving 
net helicity. Since then it is possible that one sign of the helicity 
may preferentially be transferred down to unobservably small scales.}.
The left-handedness of the magnetic field could also lead to a CP 
violating signature in the cosmic microwave background 
radiation \cite{PogVacWin01}.

\acknowledgements

I am grateful to Sean Carroll, Arnab Rai Choudhuri, Mark Hindmarsh,
Phil Kronberg, Kandu Subramanian and Alex Vilenkin 
for very useful comments. This work was supported by the DoE.

\end{document}